\documentclass[figures]{epl}

\title{Multi-particle-collision dynamics: 
       Flow around a circular and a square cylinder}
\shorttitle{Cell particle dynamics: Flow around a cylinder}
\author{A. Lamura\inst{1}, G. Gompper\inst{1}, T. Ihle\inst{2}, 
        \And D. M. Kroll\inst{2} }
\shortauthor{A. Lamura \etal}
\institute{
     \inst{1} Institut f\"{u}r Festk\"{o}rperforschung,
              Forschungszentrum J\"{u}lich,\\
              D-52425 J\"{u}lich, Germany\\
     \inst{2} Supercomputing Institute, University of Minnesota,\\
              1200 Washington Ave. S., Minneapolis, MN 55415, USA 
}

\begin{document}

\maketitle

\begin{abstract}
A particle-based model for mesoscopic fluid dynamics is used to simulate
steady and unsteady flows around a circular and a square cylinder in a 
two-dimensional channel for a range of Reynolds number between 10 and 
130. Numerical results for the recirculation length, the drag 
coefficient, and the Strouhal number are reported and 
compared with previous experimental measurements and computational fluid 
dynamics data. The good agreement demonstrates the potential of this 
method for the investigation of complex flows. 
\end{abstract}

\section{Introduction}
The simulation of the hydrodynamic behavior of complex fluids has attracted 
considerable attention in recent years. Several computational fluid dynamics 
approaches, such as lattice gas automata~\cite{fris86}, lattice-Boltzmann 
methods~\cite{mcna88,higu89}, and dissipative-particle dynamics~\cite{pago98}, 
have been proposed and developed in order to describe the dynamical behavior of  
these systems on mesoscopic length scales.   
The first two methods, despite of their conceptual simplicity, suffer
mainly from the lack of Galilean invariance. Moreover, since these are 
lattice based methods, practical applications involving irregular geometries 
often require the use of adapted computational meshes. Dissipative particle 
dynamics is an off-lattice, particle-based method which does not have these 
problems; however, it is often complex and difficult to analyze analytically.

In this Letter, we investigate in detail another particle-based 
simulation method which is a modification of Bird's Direct Simulation Monte 
Carlo algorithm \cite{bird76}. The fluid is modeled by ``particles'' whose 
positions and velocities are continuous variables, while time is discretized. 
The system is coarse-grained into the cells of a regular lattice with no 
restriction on the number of particles per cell.
The evolution of the system occurs in two steps: propagation and collision.
Each particle is first streamed by its displacement during the time 
interval $\Delta t$.
A novel algorithm for the collision step was introduced recently by 
Malevanets and Kapral~\cite{male99}. 
The cells are the collision volumes for many-particle collisions
--- which is why we refer to this method as multi-particle collision 
dynamics (MPCD) ---, which conserve mass, momentum and energy.
In Ref.~\cite{male99} it was shown that an H-theorem holds for this 
dynamics and that the correct hydrodynamic equations for an ideal gas are 
recovered. This model has been carefully studied in Ref.~\cite{ihle01}, 
where the Galilean invariance is critically discussed and it is shown 
that the original algorithm has to be modified at low temperatures to 
guarantee this symmetry.

We present the results of a quantitative analysis of the 
MPCD method in order to determine whether it can provide a convenient 
alternative to other computational fluid dynamics approaches. In particular, 
we study the problem of two-dimensional incompressible 
flow around a circular and a square cylinder. The available 
literature on this classical flow problem allows us to test and
validate this new model.

After describing the method, we present computational results for a wide 
range of Reynolds number, covering both steady and unsteady periodic flows. 
The results are compared with previous experimental and numerical studies. 

\section{The model}
The system we study consists of $N$ particles of unit mass with continuous 
positions $\vect{r}_i(t)$ and velocities $\vect{v}_i(t)$ in two dimensions, 
$i=1, 2, ..., N$.  The evolution is given by repeated streaming and collision 
steps. Assuming a unit time interval, $\Delta t = 1$, the particle positions 
change according to
\begin{equation}
\vect{r}_i(t+1) = \vect{r}_i(t) + \vect{v}_i(t) 
\label{streaming}
\end{equation}
during the streaming step. 
For the collision step, the system is divided into the cells of a regular 
lattice of mesh size $a_0$, with $a_0=1$. Each of these cells is the collision 
volume for a multi-particle collision defined by
\begin{equation}
\vect{v}_i(t+1) = \vect{u}(t) + \mathsf{\Omega} \; [\vect{v}_i(t) - \vect{u}(t)] , 
\label{collision}
\end{equation}
where $\vect{u}$ is the macroscopic velocity, defined as the average velocity 
of the colliding particles, which we assume to have coordinates of the center 
of the cell.  $\mathsf{\Omega}$ denotes a stochastic rotation matrix which 
rotates by an angle of either $+\pi/2$ or $-\pi/2$ with probability $1/2$. 
The collisions are simultaneously performed on all the particles in a cell 
with the same rotation $\mathsf{\Omega}$, but $\mathsf{\Omega}$ may differ 
from cell to cell. The local momentum and kinetic energy
do not change under this dynamics.

The division of space into a fixed cell lattice for the collision step 
breaks Galilean invariance. This is a minor effect as long as the 
mean-free path is comparable or larger than the cell size $a_0$, because 
subsequent
collisions of a particle typically occur in different cells. On the 
other hand, when the mean-free path is small compared to the cell size $a_0$,
particles remain within the same cell for many collisions, and become 
correlated. These correlations
depend on the presence of flow, and Galilean invariance is broken. 
It was shown in Ref.~\cite{ihle01} that this problem can be cured by
performing the collision operation in a cell grid which is 
shifted each time step by a random vector with components in the interval 
$[-a_0/2,a_0/2]$.
The collision environment is then independent of the macroscopic velocity, 
and Galilean invariance is exactly restored. 

\section{No-slip boundary conditions}
For planar walls which coincide with the boundaries of the
collision cells, no-slip boundary conditions are conveniently simulated 
in the case of a fixed cell lattice 
by employing a bounce-back rule, i.e. the velocities of particles which
hit the wall are inverted after the collision. However, the walls will 
generally not coincide with, or even be parallel to, the cell boundaries.
Furthermore, for small mean-free paths, when a shift operation of 
the cell lattice is required to guarantee Galilean invariance, 
partially occupied boundary cells are unavoidable, even in the simplest
flow geometries. 

\begin{figure}[ht]
\caption{Poiseuille flow through a channel of size $H=30a_0$ and 
$L=50a_0$, for $n_{av}=35$.  
(a) $k_BT=0.01275$, corresponding to a mean-free path 
$\ell = \Delta t (k_BT/m)^{1/2} = 0.11$ (with $\Delta t=m=1$). 
The simulation is
carried out with the shift operation of the collision cells. Results are
shown for simple bounce-back boundary conditions (open circles) and for 
the boundary condition (\ref{newcoll}) (full circles). The dashed and solid 
lines are fits to a parabolic flow profile. 
(b) $k_BT=0.4$, corresponding to a mean-free path $\ell = 0.63$. 
The simulation is
carried out for a fixed cell lattice, which is displaced with respect to
the walls by a quarter of the lattice constant, so that the cell lattice
fills the channel asymmetrically. 
}
\label{fig_poiseuille}
\end{figure}

We found that the simple bounce-back rule fails to guarantee no-slip boundary 
conditions in the case of partially filled cells. This can be demonstrated, 
for example, for 
a simple Poiseuille flow at low temperatures (small mean-free paths). The 
velocity profile for a channel of height $H$ and length $L$ (measured in 
units of the lattice constant) is shown in Fig.~\ref{fig_poiseuille}(a).
The velocity profile does not extrapolate to zero at the walls, and 
a strong slip is clearly visible.  
We therefore need a generalization of the bounce-back rule for partially
filled cells. Many different schemes are possible. We propose the 
following algorithm, which we found both to be very efficient and to give 
good results --- as discussed below.  
For all the cells of the channel which are cut by walls and therefore 
have a number of particles $n$ smaller than the average number $n_{av}$
of the bulk cells, we fill the  
`wall' part of the cell with virtual particles in order to make the 
effective density of real plus virtual particles equal the 
average density. The velocities of the wall particles are drawn from a
Maxwell-Boltzmann distribution of zero average velocity and the same 
temperature $T$ as the fluid. The collision step, Eq.~(\ref{collision}),
is then carried out with the average velocity of {\em all} particles
in the cell. Since the sum of random vectors drawn from a Gaussian
distribution is again Gaussian-distributed, the velocities of the 
individual wall particles never have to be determined explicitly. Instead,
the average velocity $\vect{u}$ in Eq.~(\ref{collision}) can be written
as 
\begin{equation}
\vect{u} = \frac{\sum_{i=1}^{n} \vect{v}_i + \vect{a}}{n_{av}}
\label{newcoll}
\end{equation}
where $\vect{a}$ is a vector whose components are numbers from a 
Maxwell-Boltzmann distribution with zero average and variance 
$(n_{av}-n) k_B T$ \cite{note}. Results for Poiseuille flow
with partially filled cells, both with cell-shifting and  
a fixed cell lattice, are shown in Fig.~\ref{fig_poiseuille}(a) 
and \ref{fig_poiseuille}(b), respectively. The results are in 
very good agreement with the expected parabolic flow profile
for kinematic viscosities $\nu=0.079\pm 0.001$ and $\nu=0.087\pm 0.001$, 
respectively, 
which should be compared with the values of $\nu=0.083\pm 0.001$ and 
$\nu=0.088\pm 0.001$ obtained from the (short-time) decay of the 
velocity autocorrelation function in an equilibrium system \cite{ihle01}. 
It is important to notice that the center-of-mass velocity of the virtual 
wall particles is {\em non-zero} in general, which reflects the non-zero 
temperature of the wall. Simulations of Poiseuille flow using the same 
boundary condition but with resting wall particles, i.e. 
$\vect{a}\equiv 0$ in Eq.~(\ref{newcoll}), again fails to reproduce the
desired flow profile. In this case, also the density of (real) particles 
increases drastically near the walls.

\section{Results and discussion}

\begin{figure}[ht]
\caption{Velocity field at the final steady state for $Re=30$ for the circular
(left panel) and square (center panel) cylinder. Velocity field at $Re=100$
for the square cylinder (right panel). Only a fraction of the simulation box is
shown in each case.}
\label{fig_flow}
\end{figure}

The flow around a circular and a square cylinder with diameter $D$ inside a 
plane channel of height $H$ and length $L$ was investigated. The blockage 
ratio $B=D/H$ is fixed at $0.125$ for both the cylinders. In order to reduce 
the influence of inflow and outflow boundary conditions, the length $L$ is 
set to $L/D=50$. The cylinder is centered inside the channel at a 
distance $L/4$ from the inflow region. The average number of
particles per cell is $n_{av}=10$ and the temperature is fixed at 
$k_B T=0.4$; the corresponding viscosity is $\nu=0.110 \pm 0.004$.  The 
flow is driven by assigning Maxwell-Boltzmann-distributed 
velocities with parabolic profile $v_x(y) = 4 v_{max} (H-y)y/H^2$ to 
particles in the region $0 < x \leq 10$. The maximum flow velocity, $v_{max}$, 
is fixed by the condition that the Mach number --- i.e. the velocity
relative to the speed of sound $c$, with $c^2 = [C_p/C_v] dp/d\rho = 
2k_BT/m$ --- is approximately 
$1/4$, in order to avoid significant compressibility effects. 
Periodic boundary conditions are imposed in the $x$-direction, no-slip 
boundary conditions on the channel and cylinder walls. 

Reynolds numbers, $Re\equiv v_{max} D/\nu$, in the range 
$10 \leq Re \leq 130$ were investigated. Both steady flows, with a 
closed steady 
recirculation region consisting of two symmetric vortices behind 
the body, and unsteady flows, with the well-known von Karman vortex street 
with periodic vortex shedding from the cylinder, are observed to occur for 
this range of Reynolds numbers. The critical Reynolds number above which 
flow becomes unsteady, is approximately $49$ for the circular 
cylinder~\cite{will96} and $60$ for the square cylinder~\cite{breu00}.
Figure~\ref{fig_flow} shows the macroscopic velocity field of the final 
steady-state for $Re=30$, and of the periodic vortex shedding for $Re=100$.

\begin{figure}[ht]
\caption{
The recirculation length as a function of the Reynolds number.
Square cylinder: ($\bullet$) this study with fixed cell lattice, 
($\circ$) this study with cell-shifting, ($-\!\!\!-\!\!\!-\!\!\!-$) 
Breuer {\it et al.}~\cite{breu00};
Circular cylinder: ($\star$) this study with fixed cell lattice, 
($\triangle$) this study with cell-shifting, ($- - -$) Coutanceau and 
Bouard~\cite{cout77}.}
\label{fig_recirc}
\caption{The drag coefficient $C_d = 2F_\parallel/(n_{av} v_{max}^2 D)$ --- 
where $F_\parallel$ is the force exerted on the cylinder in the direction 
parallel to the flow --- as a function of the Reynolds number.
Square cylinder: ($\bullet$) this study, ($-\!\!\!-\!\!\!-\!\!\!-$)
Breuer {\it et al.}~\cite{breu00};
Circular cylinder: ($\star$) this study, ($\circ$) Tritton~\cite{trit59},
($\triangle$) He and Doolen~\cite{he97}.}
\label{fig_drag}
\end{figure}

Such flow patterns are consistent with experiments ({\it cfr.} Fig.~4 of 
Ref.~\cite{cout77} for the circular cylinder) and simulations ({\it cfr.} 
Fig.~2 of Ref.~\cite{breu00} for the square cylinder). The length, $L_r$, 
of the recirculation region, from the rear-most
point of the cylinder to the end of the wake, has been measured in units of 
$D$.  Results, with error bars, are shown in Fig.~\ref{fig_recirc} for both 
the square and circular cylinders. The errors   
are estimated to be $1/(2D)$, since the position 
of the macroscopic velocity $\vect{u}$ in each cell is arbitrarily taken to 
be the center of the cell. Our results for $L_r/D$ are consistent with a
linear increase with Reynolds number. 
Our data for the circular cylinder compare very well with  
experimental measurements for a similar value of the blockage ratio 
($B=0.12$)~\cite{cout77}. Indeed, $L_r/D$ is a sensitive function on $B$, and 
increases with decreasing $B$~\cite{cout77}. No numerical data are 
available for a similar 
value of $B$. In the case of the square cylinder, our data can be compared
with the numerical results of Ref.~\cite{breu00}, obtained using 
lattice-Boltzmann and finite-volume methods with the same blockage ratio 
$B=0.125$ and same parabolic inflow velocity profile as in our study. A 
linear fit to our data obtained both with cell-shifting 
as well as the fixed-cell lattice data for $Re=10$ gives
\begin{equation}
L_r/D = - a + b \cdot Re
\end{equation}
with $a=0.148 \pm 0.069$ and $b=0.0525 \pm 0.0020$
while Breuer {\it et al.} find $a=0.065$ and $b=0.0554$. The slopes
are very similar, but our data are shifted to somewhat smaller values
of $L_r/D$. For the square cylinder, no experimental 
data could be found in the literature.  

Figure~\ref{fig_drag} shows the drag coefficient $C_d$ as a function of 
the Reynolds number. For the square cylinder, agreement between our 
data and the previous  
numerical measurements is satisfactory, with a small but systematic
deviation to larger values compared to Ref.~\cite{breu00} 
for $Re > 20$. In the case of the circular 
cylinder, our results are compared with the experimental measurements of 
Tritton~\cite{trit59}, and with numerical simulations by He and
Doolen~\cite{he97} performed with a lattice-Boltzmann method implemented 
on a variable mesh. In contrast to the current simulations, 
Refs.~\cite{trit59} and \cite{he97} both used a constant incoming 
velocity profile and a very small blockage ratio. 

Our results fall below (about $5\%$) the literature data in this case. 
This is mainly due to the different boundary conditions and blockage 
ratios used in the various studies. Numerical investigations with the 
lattice-Boltzmann method \cite{wagn94} showed 
that $C_d$ depends on the profile of the incoming flow. In Ref.~\cite{he97},  
lateral periodic boundaries and a constant inflow velocity profile  
instead of no-slip boundary conditions with a parabolic inflow 
velocity profile were used. It was
shown in Ref.~\cite{wagn94} that $C_d(no-slip)/C_d(periodic) = 0.82$ for 
$B=0.1$ (no error bar given). It seems reasonable to expect a similar behavior 
also for the present case. Indeed, the comparison of our results with those of 
the lattice-Boltzmann simulation \cite{wagn94} gives a ratio of 
$C_d(no-slip)/C_d(periodic) = 0.90\pm 0.03$. 

\begin{figure}[ht]
\caption{The Strouhal number as a function of the Reynolds number.
Square cylinder: ($\bullet$) this study, ($\diamond$)
Breuer {\it et al.}~\cite{breu00};
Circular cylinder: ($\star$) this study, ($\circ$) Tritton~\cite{trit59},
($\triangle$) He and Doolen~\cite{he97}, ($\ast$) Williamson~\cite{will88}.
Both lines are guides to the eye.}
\label{fig_str}
\end{figure}

In the unsteady flow regime we computed the lift coefficient 
$C_l= 2F_\perp/(n_{av} v_{max}^2 D)$, where $F_\perp$ is the force excerted
on the cylinder in the direction perpendicular to the flow. 
The characteristic frequency $f$ of the vortex shedding was determined by a 
spectral analysis (fast Fourier transform) of the time
series of $C_l$ and used to determine the Strouhal number 
$St\equiv fD/v_{max}$~\cite{note}. 
Our data as a function of $Re$ are shown in Fig.~\ref{fig_str}. 
They lie within the range of scatter of the published data.
We want to mention again that
for the circular cylinder, the existing literature were all
obtained using different inflow conditions than in our study. 

\section{Conclusions}
The multi-particle collision dynamics method has been successfully applied 
to simulate two-dimensional flow around a circular and a square cylinder. 
The results are found to be in good agreement with previous experimental 
measurements and computational fluid dynamics studies. 
The computational efficiency of the MPCD method can be estimated from
the time requirement of about 1.7 seconds (1.2 seconds) per time step for 
a system of 2 million particles with (without) cell-shifting on a 
Compaq XP1000 (667 MHz) workstation. 

The main difference between multi-particle collision dynamics and lattice
Boltzmann and finite-volume methods, which have been employed in most other
studies of this flow geometry, is the presence of thermal fluctuations.
In equilibrium, these fluctuations are responsible for a logarithmic 
divergence of the viscosity with time \cite{erns70,ihle01}, a divergence 
which is cut off by the finite system size at long times. A thermal 
renormalization of the viscosity could indeed be responsible for the 
small deviations of the values
for the recirculation length and the drag coefficient compared to the 
lattice Boltzmann results. 

The present model provides a simple alternative scheme that can be used to 
treat a wide class of physical and chemical problems. Many directions are 
accessible to exploration. The results 
presented here are relevant for studies of the 
interactions among large colloidal particles in solution. The algorithm 
can also be used for a mesoscopic model of the solvent dynamics which 
can be coupled to a microscopic treatment of solute particles. 

\acknowledgments
A.L. thanks the Supercomputing Institute of the University of Minnesota for its 
hospitality. Support from the National Science Foundation under Grant Nos.
DMR-9712134 and DMR-0083219, and the donors of the Petroleum
Research Fund, administered by the ACS, are gratefully acknowledged.

\end{document}